\documentclass[10pt, conference]{IEEEtran}
\IEEEoverridecommandlockouts

\usepackage{cite}
\usepackage{amsmath,amssymb,amsfonts}
\usepackage{algorithmic}
\usepackage{graphicx}
\usepackage{textcomp}
\usepackage{xcolor}
\usepackage[ruled,vlined]{algorithm2e}
\usepackage{textcomp}
\usepackage{physics}
\usepackage{commath}
\usepackage{subfig}
\usepackage{fancyhdr}
\def\BibTeX{{\rm B\kern-.05em{\sc i\kern-.025em b}\kern-.08em
    T\kern-.1667em\lower.7ex\hbox{E}\kern-.125emX}}

\fancypagestyle{FirstPage}{
\chead{}
\cfoot{978-1-6654-6738-4/22/\$31.00 ©2022 IEEE} 
}

\begin{document}

\title{FedREP: Towards Horizontal Federated Load Forecasting for Retail Energy Providers\\
\thanks{This work is the results of the research project funded by the Faculty of Science, Engineering and Built Environment (SEBE), Deakin University under the scheme `mini ARC analogue program (MAAP)'.}
}

\author{\IEEEauthorblockN{Muhammad Akbar Husnoo}

\IEEEauthorblockA{\textit{Centre for Cyber Security } \\
\textit{Research and Innovation (CSRI)} \\
\textit{Deakin University}\\
Geelong, VIC 3216, Australia \\
mahusnoo@deakin.edu.au}
\and
\IEEEauthorblockN{Adnan Anwar}
\IEEEauthorblockA{\textit{Centre for Cyber Security } \\
\textit{Research and Innovation (CSRI)} \\
\textit{Deakin University}\\
Geelong, VIC 3216, Australia \\
adnan.anwar@ deakin.edu.au}
\and
\IEEEauthorblockN{Nasser Hosseinzadeh}
\IEEEauthorblockA{\textit{Centre for Smart Power and } \\
\textit{Energy Research (CSPER)} \\
\textit{Deakin University}\\
Geelong, VIC 3216, Australia \\
nasser.hosseinzadeh@deakin.edu.au}
\and
\IEEEauthorblockN{Shama Naz Islam}
\IEEEauthorblockA{\textit{Centre for Smart Power and } \\
\textit{Energy Research (CSPER)} \\
\textit{Deakin University}\\
Geelong, VIC 3216, Australia \\
shama.i@deakin.edu.au}
\and
\IEEEauthorblockN{Abdun Naser Mahmood}
\IEEEauthorblockA{\textit{Department of Computer Science \& IT} \\
\textit{Latrobe University}\\
Bundoora, VIC 3086, Australia \\
A.Mahmood@latrobe.edu.au}
\and
\IEEEauthorblockN{Robin Doss}
\IEEEauthorblockA{\textit{Centre for Cyber Security } \\
\textit{Research and Innovation (CSRI)} \\
\textit{Deakin University}\\
Geelong, VIC 3216, Australia \\
robin.doss@deakin.edu.au}
}
\maketitle

\begin{abstract}

As Smart Meters are collecting and transmitting household energy consumption data to Retail Energy Providers (REP), the main challenge is to ensure the effective use of fine-grained consumer data while ensuring \textit{data privacy}. In this manuscript, we tackle this challenge for energy load consumption forecasting in regards to REPs which is essential to energy demand management, load balancing and infrastructure planning. Specifically, we note that existing energy load forecasting is centralized, which are not scalable and most importantly, is vulnerable to data privacy threats. Besides, REPs are individual market participants and liable to ensure the privacy of their own customers. To address this issue, we propose a novel horizontal privacy-preserving federated learning framework for REPs energy load forecasting, namely FedREP. We consider a differential privacy-based federated learning system consisting of a control centre and multiple retailers by enabling multiple REPs to build a common, robust machine learning model without sharing data, thus addressing critical issues such as data privacy, data security and scalability. For forecasting, we use a state-of-the-art Long Short-Term Memory (LSTM) neural network due to its  ability to learn long term sequences of observations and promises of higher accuracy with time-series data while solving the vanishing gradient problem. Finally, we conduct extensive data-driven experiments using a real energy consumption dataset. Experimental results demonstrate that our proposed federated learning framework can achieve sufficient performance in terms of MSE ranging between 0.3 to 0.4 and is relatively similar to that of a centralized approach while preserves privacy and improves scalability.

\end{abstract}

\begin{IEEEkeywords}
Smart grid, energy internet, load forecasting, federated learning, privacy, neural network, LSTM, differential privacy
\end{IEEEkeywords}

\section{Introduction}
\thispagestyle{FirstPage}
Smart grid based energy management has ensured reliable, energy-efficient, and high-quality power transfer as well as enabled end-user renewable energy integration and control. Despite the salient benefits of the smart grid, the integration of new generation of technologies including cloud computing, Artificial Intelligence, Internet of Things (IoT), etc., have introduced several challenges \cite{Fan_Ren_Feng_Liu_Wang_Lin_2021}. In particular, one of the key concerns is how to make the appropriate use of consumers' fine grained energy consumption data while ensuring \textit{data privacy}. Throughout this manuscript, we address this critical challenge related to the privacy considering a lesser studied area of distribution systems focusing on \textit{privacy preserving load forecasting for Retail Energy Providers} (REPs). 

In a traditional setting, energy load forecasting occurs at the control centre by executing advanced deep learning models which indeed achieve good forecasting performance \cite{ Gholizadeh_Musilek_2022}. However, the existing centralized approach requires retailers to share and transmit energy consumption information to the control centre where models are trained and executed. The centralized aggregation of energy consumption data faces \textbf{two} considerable challenges of privacy and security \cite{9594795, husnoo2021false} concerning such data due to sense and correlate granular data. The energy consumption data is granular enough such that one can extract individual customer’s behaviour.

Privacy preservation has been a major challenge in the roll-out of smart meters in several countries \cite{6555824}. Several studies \cite{10.1145/1878431.1878446, 10.1145/2528282.2528295} have highlighted that simple analysis of load consumption patterns recorded by smart meters can reveal household occupancy rates, the presence of people within a house, and sleep/wake-up time of residents, without any prior knowledge. Evidently, higher resolution of smart meter data leads to higher granularity in information and allows third parties to infer more sensitive information about households \cite{Farokhi_2020}. According to the Australian Privacy Foundation \cite{AustralianPrivacyFoundation}, though the initial use of sensitive energy consumption data is allowed for energy supply efficiency improvements, however, due to the sensitivity of such data, corporate and privatized energy corporations may attempt to secretly sell such data to third party organizations which then exposes serious privacy risks to consumers \cite{AustralianPrivacyFoundation}. 

To address the aforementioned issues, we propose a novel horizontal distributed energy load forecasting approach for REPs, which has competitive performance and can protect consumer's privacy. The primary contributions of this work are three-fold:

\begin{itemize}
    \item First, we have surveyed through existing studies to find related works on retail energy forecasting and have concluded that there has been very limited work on it. One possible reason behind this is REP load forecasting is technically not much different to the traditional control centre load forecasting. However, it will be an interesting topic of research to investigate how distributed learning works considering a federated setup for REP load forecasting. We investigated and found that, to the best of our knowledge, no previous work has addressed the issue of federated retail energy forecasting. This signifies the lack of research and therefore leads us to contribute to one of the earliest work within this emerging research topic.
    
    \item Second, we propose a novel load forecasting framework for REPs, namely FedREP. This framework consists of several major features: (1) Electricity consumption data will be stored at REPs and will not require sharing, (2) All REPs will collaboratively train the same model using a recent advanced distributed machine learning paradigm namely, Federated Learning, (3) Model updates will be differentially private to safeguard against leakages and (4) REPs will use the final collaboratively trained model for load forecasting purposes.
    
    \item Third, we extensively perform data-driven experiments on our proposed FedREP framework using a real energy consumption dataset namely, \textit{Solar Home Electricity Data} from Ausgrid \cite{datasetausgrid}. The experimental validations prove that our proposed framework results in good performance in terms of MSE, while ensuring privacy-preserving load forecasting at REPs. Moreover, our work performs equally well as the centralized approach as shown in Section \ref{Sec:CompaCentral}.
\end{itemize}

The rest of the paper is structured as follows. Section \ref{Sec: survey} surveys through the recent works on Load Forecasting at REPs and Federated Learning schemes. Section \ref{sec:probform} introduces the problem formulation of federated learning while Section \ref{sec:proposedframework} is focussed on the proposed FedREP framework. Section \ref{sec:results} conducts scenarios and make comparisons to verify the effectiveness of our proposed framework. Lastly, Section \ref{sec:conclusion} concludes the article and and points out the future prospects of this work.

\section{Survey of Related Work}
\label{Sec: survey}
In this section, we briefly review recent studies, focusing on energy load forecasting at REPs and federated learning schemes in smart grid.

\subsection{Load Forecasting at REPs}

Since load forecasting is an important aspect of smart grids, a multitude of recent studies \cite{Zhang_Chen_Yan_Zhang_Xu_2021, Aslam_Herodotou_Mohsin_Javaid_Ashraf_Aslam_2021, 9624274} based on several deep learning approaches have been conducted. However, most of these works focus on residential load forecasting. Hsu and Chen \cite{Hsu_Chen_2003} first proposed a machine learning solution to regional load forecasting involving REPs. They utilized a multi-layer back propagation neural network for four Taiwan regions (Northern, Eastern, Central and Southern Taiwan) load forecasting aggregated by retailers. Their centralized approach resulted in good forecasting results. A very limited number of studies \cite{Glavan_Gradi_2019} have been conducted in regards to energy load forecasting between control centres and REPs. 

As there is not much difference between REP load forecasting and traditional  control centre load forecasting, there has not been many studies on this topic. However, the lack of studies in regards to REP load forecasting opens up several opportunities for research.

\subsection{Federated Learning Schemes}

In the above literature, most of the load forecasting approaches are focused on developing and/or improving state-of-the-art models for load forecasting while being centralized and therefore, do not take into account of privacy sensitive energy consumption data. To address this issue, we identified federated learning as a viable solution to privacy-preserving data mining, as distributed nodes store and process data locally, and they can collaboratively train a distributed machine learning model by sharing model parameters with other nodes \cite{9084352}.

Despite several benefits of federated learning, it has not been fully explored and applied to load forecasting in smart grids. The earliest work in this area by Taik and Cherkaoui \cite{9148937} applied a federated approach to short-term load forecasting for residential houses and evaluated their framework with data from 200 houses from Texas, USA. Similarly, other studies \cite{Fekri_Grolinger_Mir_2021, Gholizadeh_Musilek_2022} have been working towards improving distributed short-term energy forecasting and improving privacy measures in relation to federated learning for load forecasting at customer level. However, it is worth noting that the feasibility of the proposed approach is highly dependent on the capabilities of the edge devices to perform local training. In this view, we conclude that the federated learning at consumer level for short-term load prediction is practically challenging due to low processing capabilities of current smart meters. Nonetheless, incorporating new IoT edge devices with sufficient computing requirements, will improve their practical applications.

In brief, although there exists some recent works that utilize federated learning for short-term load forecasting, no previous work has focused on load forecasting at the distributor or REPs level. Furthermore, the heterogeneity of smart meters leads to heterogeneous data  which is a major challenge in load forecasting at REPs \cite{5195790}. Therefore, to advance the state-of-the-art, we propose and develop a comprehensive federated learning framework for REP load forecasting, which is detailed and evaluated in the below sections.

\section{Problem Formulation}
\label{sec:probform}
As mentioned earlier, this manuscript aims at developing a reliable framework for multiple REPs to collaboratively train a model. Therefore, in this section, we concretely formulate the federated machine learning problem. For the horizontal separation of data, given the historical values of {$Y_1$, $Y_2$, ..., $Y_N$} of N homogeneous REPs {$X_1$, $X_2$, ..., $X_N$}, a constructed model $F^H$ is expected to learn a nonlinear mapping function by using the history-driven shared sequence feature $\textbf{x}$ to obtain the predicted value $y_i$ using the following formulation:

\begin{equation}
    y_i = F^H_{X_1, X_2, ..., X_N, Y_1, Y_2, ..., Y_N} (\textbf{x})
\end{equation}

For each REP, the history-driven shared sequence feature values x are consistent and transparent to all REPs as they share the same feature set. During the training process, raw training data {$X_1$, $X_2$, ..., $X_N$} and {$Y_1$, $Y_2$, ..., $Y_N$} will at no circumstance be shared to other REPs. All intermediate results (such as model parameters e.g. [[$w^n_k$]], [[$w_0$]]) are expected to be properly encrypted before they are transferred.

Furthermore, since training samples are held by several REPs where local models are trained and aggregated by the control centre, we formulate the aggregation of local models as an optimization problem in the form of:

\begin{equation}
    F^H(\textbf{x}) = \sum^H_{h=1} \dfrac{n_h}{n} F_h(\textbf{x})
\end{equation}

\noindent
whereby h is the number of REPs, $n$ is the number of training samples and $F_h$ is the local model.

\section{Proposed FedREP Framework}
\label{sec:proposedframework}
Throughout this section, we will present the main ideas and components of the proposed FedREP framework. In particular, we will firstly introduce the system model followed by the Federated Learning infrastructure.

\subsection{System Model}

\begin{figure}
    \centering
    \includegraphics[width=6cm]{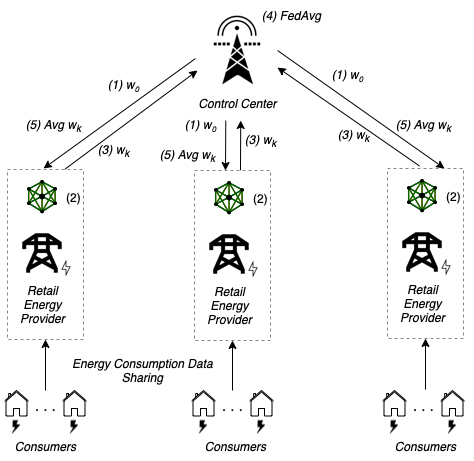}
    \caption{An illustration of FedREP framework}
    \label{fig:FedREPillustration}
\end{figure}

As previously discussed, the objective of this study is to design a federated learning framework, FedREP, that collaboratively trains a forecasting model for REP load forecasting in a distributed manner. As shown in Fig \ref{fig:FedREPillustration}, the proposed FedREP framework consists of three components as discussed.

\begin{enumerate}
    \item \textit{Smart Meter}: Each customer has a smart meter that is connected to one of the REPs. Each smart meter collects energy consumption data and forwards the data to its respective REP.
    \item \textit{Retail Energy Provider}: Retail Energy Providers (REPs) receive energy consumption data from smart meters of different houses and is the owner of these data. Common assumptions include sufficient storage and computation capabilities to store historical energy data and locally train a forecasting model. 
    \item \textit{Control Centre}: The control centre is responsible for broadcasting a learning model and default model parameters, aggregation of parameters after training and finally broadcasting the updated model parameters.
\end{enumerate}

\subsection{Proposed Federated Learning Approach}

\begin{algorithm}
\textit{Control Centre} distributes the unanimous model $F$ and encrypted parameter initialization [[$w_0$]] to all \textit{REPs} 1,2,...,N.

\For{each communication round, $k = 1, 2,...,K$}{

\For{each \textit{REP}, $n = 1,2,...,N$}{

Decrypt [[$w_{k-1}$ = $Dec([[w_{k-1}]])$]].

Update $w_{k-1}$ by training local dataset $w^n_k$:= $w_{k-1}$-$\eta g^n_k$ where $\eta$ is the learning rate and $g^n_k$ denotes the calculated gradient in training epoch.

Perturb $w^n_k$ with sufficient noise such that $\mathcal{M}w^n_k$ is differentially private.

Encrypt $\mathcal{M}w^n_k$ into [[$\mathcal{M}w^n_k$]] and send to control centre.

Calculate MSE loss $L^n_k$ and send it to server.
}

\textbf{end}

\textit{Control Centre} averages the encrypted parameters (in the cyphertext space) [[$w_k$]] :=
$\overline{\overline{AVG}_n}$[[$w^n_k$]]. 

\textit{Control Centre} averages losses $L_k$:=$AVG_n (L^n_k)$.

\textit{Control Centre} perturbs the model updates $\mathcal{M} w_{k}$.

\textit{Control Centre} distributes $L_k$ and [[$w_k$] to all \textit{REPs}.

\textbf{If Agreed on Convergence, Terminate.}

}
\textbf{end}
\caption{FedREP Framework.}
\label{proposedalgo}
\end{algorithm}

As mentioned earlier, federated learning is a form of machine learning where most of the training process is done in a distributed way among devices referred to as clients. Therefore, an iteration of our proposed Federated Learning goes as follows: (1) Control Centre distributes unanimous model and encrypted parameter initialization to all REPs. (2) Local models are trained by each REP with their own local dataset. (3) After training, REPs initially locally perturb their model parameters and send their encrypted model parameters back to the control centre. (4) Control Centre aggregates and averages the parameters from different REPs using \textit{FederatedAveraging} algorithm as detailed in Section \ref{sec:Fedavg}. (5) Control center in turn perturbs the averaged model updates before distributing the updated parameters to the REPs for another training round. (6) This process is repeated until convergence is reached. The steps and operations have been detailed in Fig. \ref{fig:FedREPillustration} and Algorithm \ref{proposedalgo} respectively. 

\subsection{Learning Model}

As earlier mentioned, the main objective of this paper is to propose a federated learning framework for load forecasting at REPs. Since we are dealing with time-series data, we use a state-of-the-art deep learning model known as Long Short-Term Memory (LSTM) as our forecasting model due to its ability to learn long term  sequences  of  observations and promises of higher accuracy with time-series data while solving the vanishing gradient problem \cite{elsworth2020time}. Generally, LSTMs consist of five main elements namely: (1) Input Gates, (2) Forget Gates, (3) Output Gates, (4) Cells, and lastly, (5) State Gates. More specifically, our proposed FedREP framework uses a 2-layer stacked LSTM model. We set the number of hidden units to 256 and 128 respectively. In addition, we use a regular and recurrent dropout rate of 0.2 and apply ReLU activation function in order to avoid over-fitting.

\subsection{Control Centre Aggregation}
\label{sec:Fedavg}

As mentioned earlier, the control centre is responsible for the aggregation of locally trained models. In our proposed FedREP framework, we use \textit{FederatedAveraging (FedAvg)} algorithm to orchestrate the training as shown in Algorithm \ref{proposedalgofedavg}.

\begin{algorithm}
initialize $w_0$

\For{each communication round, t = 0, 1, 2, ... }{
$m$ $\longleftarrow$ $max([$C.K$], 1)$ where $C$ represents fraction of REPs to select for training and $K$ is the REPs across which data and computation are distributed.

$S_{t}$ = set of $m$ REPs.

\For{each REP n $\in S_{t}$ \textbf{in parallel}}{

$w^n_{k+1}$ = ClientUpdate(k, $w_k$)

$w_{k+1} = \sum_{k \in S_t} \dfrac{n_k}{n} w^n_{k+1}, n = \sum_{k \in S_t} n_k$
}
}
\caption{FedAvg Algorithm At Control Centre.}
\label{proposedalgofedavg}
\end{algorithm}

\section{Simulation \& Results}
\label{sec:results}
In this section, we firstly introduce the dataset and metric used during the simulation and eventually thoroughly evaluate our proposed federated approach.

\subsection{Dataset Pre-processing \& Evaluation Method}

This research was conducted using \textit{Solar Home Electricity Data} from Eastern Australia's largest electricity distributor, Ausgrid \cite{datasetausgrid}. The dataset composes of half-hourly electricity consumption data of 300 de-identified customers which is measured using gross meters during the period starting 1\textsuperscript{st} July 2012 to 30\textsuperscript{th} June 2013. We initially filter the data based on General Consumption (GC) category. It is then converted to the suitable time-series format. The data of an energy retailer is prepared by aggregating customer's data present within a selected postcode. It is then scaled between 0 and 1, and is split into train (70\%) and test (30\%) subsets. Lastly, we transform the time series into sliding windows with look-backs of size 12 and a look-ahead of size 5.

We use Mean Squared Error (MSE) loss to evaluate the LSTM model's performance with regard to the prediction error. The expression for MSE loss is as follows:

\begin{equation}
    MSE = \dfrac{1}{N}\sum_{i=1}^N (y_{i} - \hat{y_i})\textsuperscript{2}
\end{equation}

\noindent
where $\hat{y_i}$ is the predicted value, $y_i$ is the actual value and $N$ is the number of predicted values.

\subsection{Simulation Scenarios}

All simulations throughout this manuscript have been performed on Colab Pro version due to higher GPU and RAM access. The different scenarios (with increasing number of energy retailers) that were evaluated are summarized in Table \ref{tab:evaluatedscenarios}. We first set the centralized approach as a benchmark. During each federated learning scenario, we increase the number of energy retailers to see the effect of the addition of new participants on the global model. In all the federated scenarios, the communication round was set to 80 rounds.  In particular, each scenario is as follows:

\begin{enumerate}
    \item \textit{Scenario 1}: We choose to include 4 federated REPs. Each REP consists of the aggregated energy consumption of the one of the 4 postcodes, \textit{2287}, \textit{2289}, \textit{2291} and \textit{2292} of Newcastle, NSW.

    \item \textit{Scenario 2}: We choose to include 6 federated REPs. Each REP consists of the aggregated energy consumption of the one of the 6 postcodes, \textit{2287}, \textit{2289}, \textit{2291}, \textit{2292}, \textit{2293} and \textit{2294} of Newcastle, NSW.

    \item \textit{Scenario 3}: We choose to include 8 federated REPs. Each REP consists of the aggregated energy consumption of the one of the 8 postcodes,\textit{2287}, \textit{2289}, \textit{2291}, \textit{2292}, \textit{2293},  \textit{2294}, \textit{2296} and \textit{2297}  of Newcastle, NSW.
\end{enumerate}

\begin{figure*}
    \centering
    \subfloat[\centering Train Loss (4 REPs)]{{\includegraphics[width=5cm]{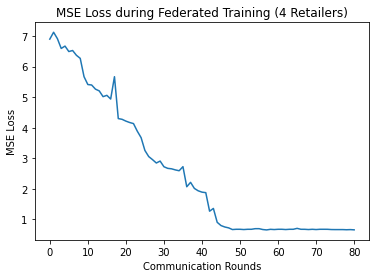} }}%
    \qquad
    \subfloat[\centering Train Loss (6 REPs)]{{\includegraphics[width=5cm]{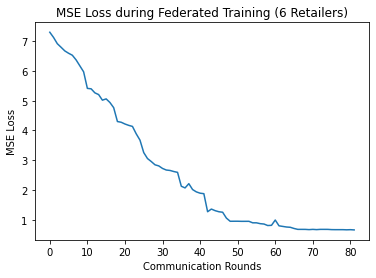} }}%
    \qquad
    \subfloat[\centering Train Loss (8 REPs)]{{\includegraphics[width=5cm]{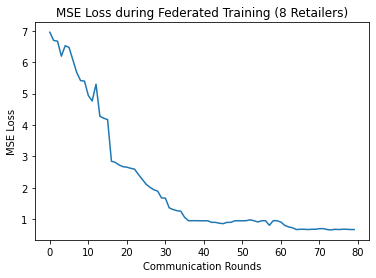} }}%
    \caption{Convergence of Federated Model based on the number of REPs.}%
    \label{fig:trainloss}
\end{figure*}

\begin{table*}[!h]
    \centering
    \caption{Evaluated Scenarios}
    \begin{tabular}{|c|c|c|c|c|c|}
    \hline
    \textbf{Scenarios} & \textbf{Number of Retailers} &\textbf{Minimum MSE} & \textbf{Maximum MSE} & \textbf{Mean MSE}\\\hline
    1 & 4 &0.328981 &0.349443 &0.338211 \\\hline
    2 & 6 &0.378698 &0.412812 &0.39321 \\\hline
    3 & 8 &0.42803 &0.4534395 &0.436892 \\\hline
    \end{tabular}
    \label{tab:evaluatedscenarios}
\end{table*}

\begin{figure*}[!h]
    \centering
    \subfloat[\centering Global Model Trained with 4 REPs data]{{\includegraphics[width=5cm]{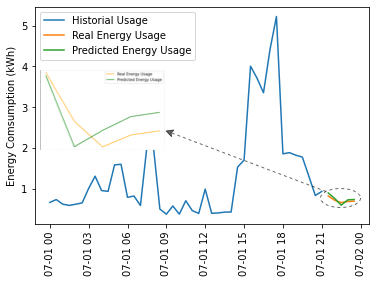} }}%
    \qquad
    \subfloat[\centering Global Model Trained with 6 REPs data]{{\includegraphics[width=5cm]{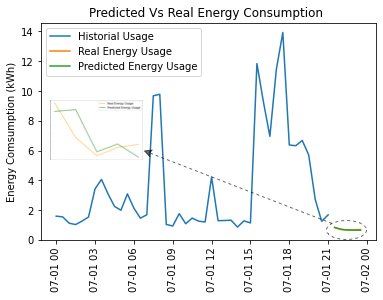} }}%
    \qquad
    \subfloat[\centering Global Model Trained with 8 REPs data]{{\includegraphics[width=5cm]{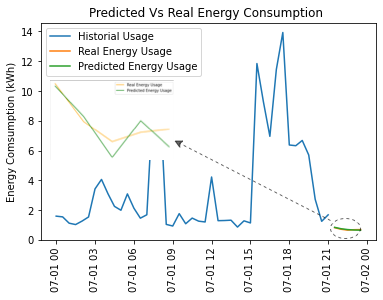} }}%
    \caption{Predictions for the next five half-hours for a retailer who did not participate in the training.}%
    \label{fig:prediction}
\end{figure*}

\subsection{Empirical Results \& Discussions}

In this section, we thoroughly discuss the evaluation results of our proposed architecture. As  mentioned earlier, the models are evaluated in terms of MSE as shown in Table \ref{tab:evaluatedscenarios}. We record the minimum, maximum and mean MSE loss for each scenario after conducting several experiments. We summarize the results based on scenarios as follows:

\begin{enumerate}
    \item \textit{Scenario 1}: The mean MSE loss recorded is 0.338211. Furthermore, from Fig. \ref{fig:prediction} (a), we can see that the global model forecasts are very close to the actual usage. This shows that the global model generalized significantly well on new data that did not participate in the training process and therefore, signifies that our proposed FedREP framework is within acceptable performance.
    \item \textit{Scenario 2}: The mean MSE loss recorded for this scenario is relatively consistent with Scenario 1. Judging from Fig. \ref{fig:prediction} (b), we can similarly deduce that the global model achieves acceptable forecasting performance. However, as opposed to Scenario 1, we can see an increase in the number of communication rounds required for convergence.
    
    \item \textit{Scenario 3}: Similar to the two previous scenarios, we see a relatively consistent loss and performance of our model. In comparison with Scenario 1 \& 2, it can be seen that an increase in the number of retailers also increases the time to convergence of our proposed FedREP approach. Therefore, we can deduce that the number of REPs is directly proportional to the number of communication rounds to convergence.
    
\end{enumerate}

\begin{figure}
    \centering
    \subfloat[\centering Convergence of Centralized Model vs FedREP Validation Scenarios]{{\includegraphics[width=5cm]{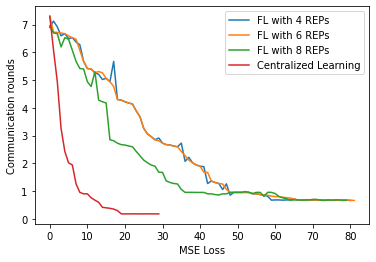} }}%
    \qquad
    \subfloat[\centering Forecasting using Centralized Model]{{\includegraphics[width=5cm]{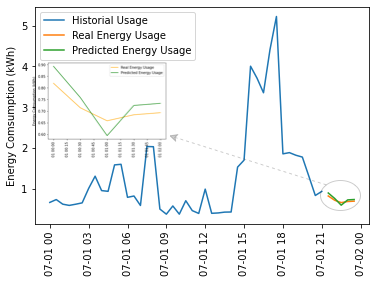} }}
    \caption{Centralized model results after training.}%
    \label{fig:predictioncentra}
\end{figure}

It is worth noting that, although the Mean Squared Error Loss stays relatively consistent with increasing number of REPs, there is a slight decrease in the performance which is due to the stochastic nature of energy load profiles. This issue can be mitigated in future works by leveraging personalisation techniques. Furthermore, as seen on the graphs in Figure \ref{fig:trainloss}, an increase in the number of results in increasing rounds to convergence, hence increasing the time to convergence of the model. Lastly, we use the global models for each scenario and test them on a new REP data that did not participate within the model training sessions. From the time series graphs present in Figure \ref{fig:prediction}, we can deduce that our proposed federated learning approach offers good forecasting performance. Lastly, as mentioned earlier, we see an increase in the communication rounds to convergence with increasing number of REPs and therefore, we can also deduce that the number of REPs is directly proportional to the number of communication rounds required for convergence of the global model during training. 

\subsection{Comparison against Centralized Model}
\label{Sec:CompaCentral}

In addition to the three experimental federated scenarios as mentioned in Table \ref{tab:evaluatedscenarios}, we further compare our proposed FedREP framework with a centralized version. The centralized version is trained with 30 epochs. From Fig. \ref{fig:predictioncentra} (a), it can be seen that the centralized model converges faster than within our proposed federated approach. Furthermore, based on the Figure \ref{fig:predictioncentra} (b), the forecasting performance of the centralized approach and our proposed federated framework is comparatively similar which indicates the practical application of our framework to the smart grid paradigm. However, centralized model training requires the uploading of data to the control centre which, as earlier mentioned, increases risks of privacy threats. Therefore, our proposed FedREP is a privacy-preserving alternative to centralized learning within the smart grid ecosystem.

\section{Conclusion \& Future Works}
\label{sec:conclusion}
Load forecasting is a challenging task considering the stochastic nature of consumption profiles. This paper presents a federated approach for energy load forecasting for retail energy providers to tackle data diversity and, most importantly, privacy challenges in smart grids. We evaluated our proposed approach using a real-world dataset in three different scenarios (by increasing the number of federated retail energy providers). The experimental validations reveal that sufficient performance of the global model on new data. 

Although this paper studies the case of federated learning with regards to retail energy providers, we believe that our framework can have further implications on the emerging concept of Energy Internet. While it aims to interconnect all energy components within the smart grid context, our federated learning framework can further extend its power by leveraging state-of-the-art cloud-computing and edge-computing technologies. We will explore these topics in the future works.

\bibliographystyle{IEEEtran}
\bibliography{refs.bib}

\end{document}